\documentclass[11pt,a4paper]{article}

\usepackage[utf8]{inputenc}
\usepackage[T1]{fontenc}
\usepackage[english]{babel}
\usepackage{lmodern}
\usepackage{microtype}

\usepackage[
  top=25mm, bottom=25mm,
  left=25mm, right=25mm
]{geometry}

\usepackage{amsmath, amssymb}
\usepackage{graphicx}
\usepackage{float}
\usepackage{booktabs}
\usepackage{array}
\usepackage{xcolor}
\usepackage{hyperref}
\usepackage{url}
\usepackage{authblk}
\usepackage{abstract}
\usepackage{setspace}
\usepackage{caption}
\usepackage{subcaption}
\usepackage{parskip}

\hypersetup{
  colorlinks = true,
  linkcolor  = blue!60!black,
  citecolor  = blue!60!black,
  urlcolor   = blue!60!black,
  pdftitle   = {A Decade of Public Procurement in Spain},
  pdfauthor  = {Manuel Muñoz Plá}
}

\title{%
  \textbf{A Decade of Public Procurement in Spain:}\\
  \textbf{Extraction, Structuring, and Analysis}\\
  \textbf{of BOE Data (2014--2024)}
}

\author[1]{Manuel Muñoz Plá%
  \thanks{ORCID: \href{https://orcid.org/0009-0000-5714-912X}%
    {0009-0000-5714-912X}}}

\affil[1]{Independent Researcher}

\date{May 2025}

\begin{document}

\maketitle

\begin{abstract}
Public procurement data in Spain are disseminated through the Official
State Gazette (\textit{Boletín Oficial del Estado}, BOE), a
semi-structured legal publication that hinders systematic computational
analysis. This paper presents a longitudinal, open dataset of
approximately 97,000 public procurement records extracted from Section
V-A of the BOE covering the period January 2014 to December 2024. We
describe a fully reproducible ETL pipeline that transforms heterogeneous
HTML announcements into a structured dataset of 15~variables aligned
with the Open Contracting Data Standard (OCDS), integrating financial,
administrative, sectoral (CPV), and geographic attributes.

Exploratory analysis reveals strong positive skewness in contract values,
geographic concentration in major administrative regions, and sectoral
dominance of services and supplies over the decade. We apply multiple
linear regression to examine determinants of awarded contract value,
K-Means clustering to segment contractors by volume and aggregate amount,
and a Wilcoxon--Mann--Whitney non-parametric test to confirm
statistically significant differences between public works and services
contracts.

The dataset is published under a CC0 licence at
\url{https://doi.org/10.5281/zenodo.15120882}. The extraction pipeline
is publicly available at
\url{https://github.com/mmunozpl/M2.851-May25}. Together they
constitute an open, reproducible resource for public procurement
research and evidence-based policy analysis in Spain.
\end{abstract}

\noindent\textbf{Keywords:} public procurement, open government data,
BOE, Spain, OCDS, ETL pipeline, data science, contract awards, CPV.

\vspace{6pt}
\noindent\textbf{Dataset DOI:}
\href{https://doi.org/10.5281/zenodo.15120882}%
{10.5281/zenodo.15120882} (CC0)

\vspace{2pt}
\noindent\textbf{Code:}
\url{https://github.com/mmunozpl/M2.851-May25}

\hrule
\vspace{8pt}

\section{Introduction}

Public procurement constitutes a principal mechanism through which
public expenditure reaches the private sector. In the European Union it
accounts for approximately 14--20\% of GDP, representing one of the
largest categories of public expenditure~\cite{ec2023}. The
availability of structured, machine-readable procurement data is
therefore a prerequisite for empirical research on public spending
patterns, market structure, and administrative efficiency.

In Spain, the \textit{Boletín Oficial del Estado} (BOE) constitutes
the authoritative source of public contracting announcements at the
central state level. Section~V-A of the BOE publishes both tender
notices (\textit{licitaciones}) and contract award notices
(\textit{formalizaciones}) for all central-state contracting
authorities. However, the BOE publishes in HTML and PDF formats that
are semi-structured, heterogeneous across institutions and time
periods, and not designed for computational consumption. No fully
structured, machine-readable national dataset covering a longitudinal
period of central-state procurement exists in the public domain.

The objective of this work is threefold. First, to build and publish a
reproducible pipeline that transforms a decade of BOE procurement
announcements into a structured dataset. Second, to align that dataset
with the Open Contracting Data Standard (OCDS) to enable
interoperability with international procurement data~\cite{ocds2020}.
Third, to demonstrate the analytical value of the resulting dataset
through exploratory, supervised, and unsupervised analyses of
procurement patterns in Spain from 2014 to 2024.

Data extraction and reuse are explicitly permitted under Spanish
Law~37/2007 on the reuse of public sector information
(RISP)~\cite{risp2007}, which designates BOE content as open, freely
reusable public data. The dataset is published under a Creative Commons
Zero (CC0) licence, ensuring maximum accessibility in accordance with
FAIR data principles~\cite{wilkinson2016}.

The contributions of this work are:
\begin{enumerate}
  \item[(i)]   a longitudinal, OCDS-aligned dataset of Spanish
               central-state public procurement (2014--2024), published
               under CC0 at
               \url{https://doi.org/10.5281/zenodo.15120882};
  \item[(ii)]  a fully reproducible ETL pipeline for structured
               extraction from BOE HTML announcements, publicly
               available at
               \url{https://github.com/mmunozpl/M2.851-May25};
  \item[(iii)] empirical evidence on contract value distribution,
               geographic concentration, sectoral structure, and
               contractor market stratification in Spanish public
               procurement over a decade.
\end{enumerate}

\section{Related Work}

This work lies at the intersection of open government data, public
expenditure analysis, and administrative data science.

\textbf{Open contracting data.}
The Open Contracting Data Standard (OCDS) provides a free,
non-proprietary data model for publishing information across the full
contracting cycle --- planning, tendering, award, and contract
management. Implemented by over 50 governments and endorsed by the
G20, OCDS constitutes the reference standard for structured procurement
data~\cite{ocds2020}. Alignment with OCDS enables comparative research
and interoperability across national datasets.

\textbf{Structured procurement datasets.}
Several national and supranational bodies have published structured
procurement datasets. The EU's Tender Electronic Daily (TED) provides
machine-readable notices across member states. The United Kingdom
(Contracts Finder), Ukraine (ProZorro), and Colombia (SECOP) have
published OCDS-aligned open datasets enabling economic and governance
analysis. Spain lacks an equivalent structured national dataset at
central-state level; the present work addresses this gap.

\textbf{Administrative data science.}
The transformation of semi-structured administrative records into
analytical datasets is a recognised methodological challenge. Heterogeneous
formats, inconsistent entity naming, missing values arising from
administrative life cycles, and reporting biases inherent to legal
publications require domain-specific cleaning strategies~\cite{han2011,
osborne2013}.

\section{Data Source and Extraction Architecture}

\subsection{Source}

The dataset derives from \textbf{Section~V-A of the BOE}, which
publishes procurement announcements for all contracting authorities at
the Spanish central state level. The temporal coverage spans
1~January~2014 to 31~December~2024. The source is official, in the
public domain, and freely reusable under Law~37/2007 RISP~\cite{risp2007}.

\subsection{ETL Pipeline}

The extraction follows an ETL (Extract--Transform--Load) architecture
implemented in Python (3.9+) and publicly available at
\url{https://github.com/mmunozpl/M2.851-May25}. The pipeline is
modular, with clear separation of responsibilities:

\begin{itemize}
  \item \texttt{obtener\_anuncios.py} --- parses daily BOE indices and
        identifies contracting notices by section and announcement type.
  \item \texttt{obtener\_analisis.py} --- extracts contract modality,
        type, procedure, geographic scope, and CPV category descriptions.
  \item \texttt{obtener\_datos\_economicos.py} --- extracts financial
        fields (estimated value, awarded value) and contractor names from
        structured HTML blocks and free text.
  \item \texttt{obtener\_extra\_texto.py} --- extracts granular CPV codes.
  \item \texttt{main.py} --- orchestrates the complete pipeline and
        persists results to CSV.
  \item \texttt{test.py} --- unit tests for module validation.
\end{itemize}

HTTP connections use \texttt{requests.Session()} with configurable
timeouts, automatic retries, and inter-request delays, ensuring robust
and responsible scraping. Requirements are pinned in
\texttt{requirements.txt} (requests~2.31.0, BeautifulSoup4~4.12.3,
pandas~2.0.3).

\section{Dataset Construction and Variables}

\subsection{From Announcements to Records}

The initial scraping yields \textbf{90,140~observations} covering both
tender notices and contract award announcements. In a second phase,
four new variables are added by re-parsing each HTML announcement:
\textit{Naturaleza} (contract category), \textit{valor\_estimado\_
licitacion} (pre-award estimated value, €),
\textit{valor\_oferta\_adjudicada} (final awarded value, €), and
\textit{nombre\_adjudicatario} (contractor name). Records with multiple
awardees are disaggregated into one observation per contractor,
producing a raw dataset of \textbf{97,154~observations across
15~variables}.

\subsection{Variable Description}

\begin{table}[H]
\centering
\caption{Dataset variables, types, and descriptions.}
\label{tab:variables}
\small
\begin{tabular}{lll}
\toprule
\textbf{Variable} & \textbf{Type} & \textbf{Description} \\
\midrule
\texttt{Institucion}               & character & Principal contracting institution \\
\texttt{Organismo\_responsable}    & character & Managing department or unit \\
\texttt{Expediente}                & character & Unique procedure identifier \\
\texttt{Fecha}                     & Date      & Publication date \\
\texttt{Tipo}                      & character & Licitación / Contratación \\
\texttt{Naturaleza}                & character & Contract category \\
\texttt{Objeto}                    & character & Free-text contract description \\
\texttt{Procedimiento}             & character & Procurement procedure \\
\texttt{Ambito\_geografico}        & character & Geographic scope \\
\texttt{Materias\_CPV}             & character & CPV category descriptions \\
\texttt{Codigos\_CPV}              & character & Specific CPV codes \\
\texttt{valor\_estimado\_licitacion} & numeric & Pre-award estimated value (€) \\
\texttt{valor\_oferta\_adjudicada} & numeric   & Final awarded value (€) \\
\texttt{nombre\_adjudicatario}     & character & Awarded contractor name \\
\texttt{Enlace\_HTML}              & character & URL to BOE announcement \\
\bottomrule
\end{tabular}
\end{table}

\section{Data Cleaning and Preprocessing}

\subsection{Type Conversions and Normalisation}

All variables are initially read as character strings. \texttt{Fecha}
is converted to Date format. The two monetary fields are converted to
numeric, removing currency symbols, thousand separators (\texttt{.})
and decimal commas (\texttt{,}).

Categorical variables are normalised: inconsistent casing, punctuation
variants, and orthographic differences are resolved through uppercase
normalisation and entity deduplication (e.g., \texttt{"ADIF -
Presidencia"} and \texttt{"Adif Presidencia."} are unified as
\texttt{"ADIF PRESIDENCIA"}). CPV fields are reduced to numeric codes.

\subsection{Missing Values}

Table~\ref{tab:missing} summarises the principal missing-value patterns.
The complementary behaviour of the two monetary fields reflects the
administrative contracting life cycle: once awarded, the estimated value
becomes historical. No imputation is performed on monetary fields.

\begin{table}[H]
\centering
\caption{Principal missing-value patterns in the dataset.}
\label{tab:missing}
\small
\begin{tabular}{lrl}
\toprule
\textbf{Variable} & \textbf{Missing (\%)} & \textbf{Nature} \\
\midrule
\texttt{valor\_estimado\_licitacion} & 54.67 & Present in unawarded tenders only \\
\texttt{valor\_oferta\_adjudicada}   & 35.10 & Absent in unawarded tenders \\
\texttt{nombre\_adjudicatario}       & 35.10 & Tied to absence of awarded value \\
\texttt{Procedimiento}               &  5.88 & Not reported in some announcements \\
\texttt{Ambito\_geografico}          &  6.61 & Likely national-scope contracts \\
\texttt{Codigos\_CPV}                &  8.95 & Absent in older announcements \\
\bottomrule
\end{tabular}
\end{table}

\subsection{Filtering}

Records with \texttt{valor\_oferta\_adjudicada} $= 0$ are removed as
administrative artefacts without economic content. Records with awarded
value $< 1{,}000$~€ are also removed (1,475~records, 2.38\% of total),
as they correspond to extraction errors or negligible microcontracts.
After full cleaning the dataset contains \textbf{96,032~observations}.

\subsection{Analytical Subset}

For modelling and statistical analysis, only \textit{Contratación}
records (contract award notices with confirmed amounts and identified
contractors) are retained. This yields \textbf{60,456~final
observations} across 12~analytical variables, spanning 2,994~distinct
publication dates from 2014-01-03 to 2024-12-31, produced by
79~unique contracting institutions and 1,203~responsible bodies.

\section{Exploratory Analysis}

\subsection{Distribution of Contract Values}

\begin{figure}[H]
  \centering
  \includegraphics[width=0.60\textwidth]{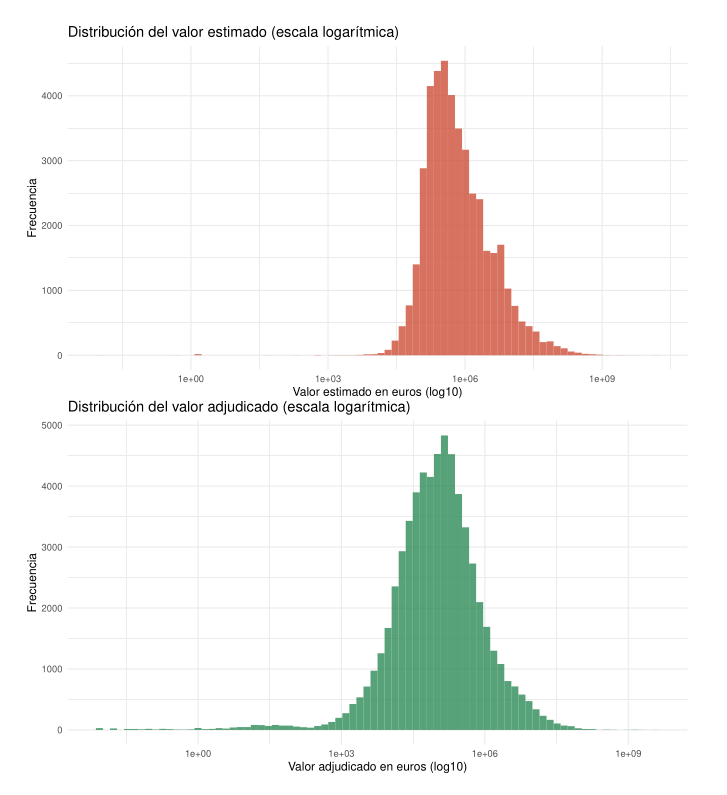}
  \caption{Distribution of awarded contract values in log10 scale.
           Most contracts concentrate between $10^5$ and $10^6$~euros,
           with long tails at both extremes.}
  \label{fig:hist}
\end{figure}

Awarded contract values exhibit strong positive skewness, spanning
approximately six orders of magnitude (Table~\ref{tab:stats}).
Figure~\ref{fig:hist} shows the distribution in log10~scale.

\begin{table}[H]
\centering
\caption{Descriptive statistics of awarded contract values.}
\label{tab:stats}
\begin{tabular}{lr}
\toprule
\textbf{Statistic} & \textbf{Value (€)} \\
\midrule
Mean    & 1,243,788 \\
Median  & 110,436   \\
Minimum & 1,000     \\
Maximum & 4,317,800,000 \\
\bottomrule
\end{tabular}
\end{table}

\begin{figure}[H]
  \centering
  \includegraphics[width=0.60\textwidth]{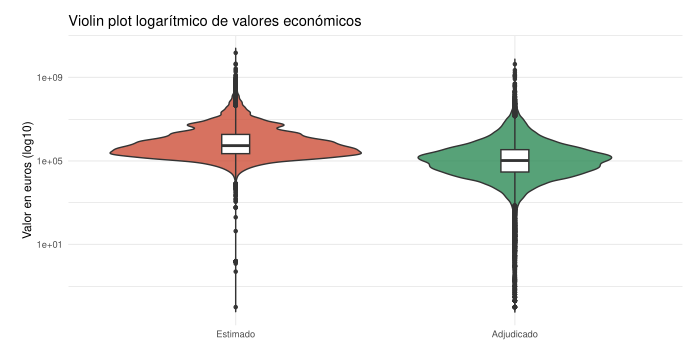}
  \caption{Logarithmic violin plot of estimated and awarded values.
           Awarded amounts track estimated values closely around the
           median, with tighter concentration and shorter tails,
           consistent with competitive adjustment during the tender
           process.}
  \label{fig:violin}
\end{figure}

A three-tier classification by quartile --- microcontracts (below Q1,
median $\approx$~12,009~€), standard contracts (Q1--Q3, median
$\approx$~104,447~€), and macrocontracts (above Q3, median
$\approx$~912,738~€) --- captures the structural heterogeneity of the
dataset.

\subsection{Geographic Distribution}

Procurement activity is markedly concentrated in a small number of
regions. Table~\ref{tab:geo} lists the top~5 geographic scopes by
contract count. Madrid's dominance reflects the administrative
concentration of central-state bodies.

\begin{table}[H]
\centering
\caption{Top~5 geographic scopes by number of contract awards.}
\label{tab:geo}
\begin{tabular}{lr}
\toprule
\textbf{Geographic scope} & \textbf{N contracts} \\
\midrule
Comunidad de Madrid & 22,138 \\
Andalucía           & 5,792  \\
Nacional            & 4,439  \\
Sin definir         & 3,782  \\
Galicia             & 2,773  \\
\bottomrule
\end{tabular}
\end{table}

\subsection{Sectoral Structure and Temporal Trend}

Services and Supplies together account for over 85\% of contract award
records throughout the decade, with Works as the third largest
category. The year-on-year increase in total contract count is
consistent with progressive digitalisation of procurement publication
and with the extraordinary expenditure cycles associated with COVID-19
emergency procurement (2020) and European Next Generation EU funds
(2021--2023).

\section{Analysis}

\subsection{Supervised Analysis: Multiple Linear Regression}

A multiple linear regression is fitted to predict
\texttt{valor\_oferta\_adjudicada} from categorical predictors:
\texttt{Institucion}, \texttt{Naturaleza}, \texttt{Ambito\_geografico},
\texttt{Procedimiento}, \texttt{Organismo\_responsable}, and calendar
year (\texttt{Ano}). The dataset is split 70\% training / 30\% test,
with factor levels harmonised across splits.

\begin{table}[H]
\centering
\caption{Performance metrics of the multiple linear regression model
         on the test set.}
\label{tab:model}
\begin{tabular}{lr}
\toprule
\textbf{Metric} & \textbf{Value} \\
\midrule
MAE          & 1,911,532~€  \\
RMSE         & 11,066,601~€ \\
$R^2$        & 0.014        \\
Adjusted $R^2$ & 0.005      \\
\bottomrule
\end{tabular}
\end{table}

The low $R^2$ is expected given the structural absence of imprescindible
predictive variables (contract duration, bidder count, lot structure).
The model captures aggregate structural effects: the largest
coefficients in absolute magnitude correspond to specific geographic
scope combinations and the contract category \textit{Gestión de
Servicios Públicos}, confirming that geographic and functional
dimensions account for the largest share of explained variance.

\begin{figure}[H]
  \centering
  \includegraphics[width=0.72\textwidth]{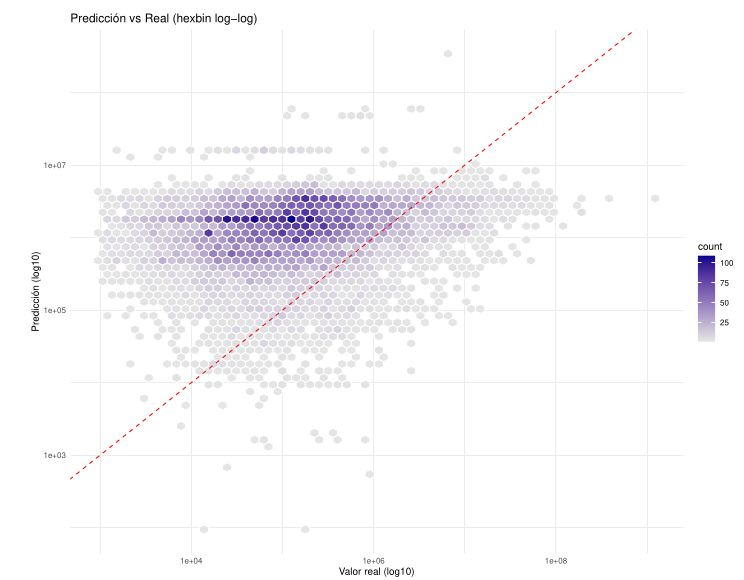}
  \caption{Hexbin scatter of predicted vs.\ real awarded values in
           log-log scale. Concentration around the diagonal denotes
           acceptable accuracy for standard contracts;
           dispersion increases at both extremes of the distribution.}
  \label{fig:hexbin}
\end{figure}

Figure~\ref{fig:hexbin} shows the distribution of prediction errors
in log-log~scale: the model achieves reasonable accuracy for standard
contracts but exhibits increasing dispersion at both tails.

\subsection{Unsupervised Analysis: K-Means Clustering of Contractors}

K-Means clustering is applied to the 16,502~unique contractors using
two log-transformed aggregated features: total contracts awarded and
total awarded value. The elbow method identifies $k = 3$ as optimal.

\begin{table}[H]
\centering
\caption{K-Means cluster summary ($k=3$).}
\label{tab:clusters}
\small
\begin{tabular}{clrrrl}
\toprule
\textbf{Cluster} & \textbf{Profile} & \textbf{N} &
\textbf{Median contracts} & \textbf{Median value (€)} \\
\midrule
1 & High-value operators & 3,710 & 3 & 4,428,634 \\
2 & Standard operators   & 7,770 & 1 & 314,049   \\
3 & Microoperators       & 5,022 & 1 & 32,258    \\
\bottomrule
\end{tabular}
\end{table}

\begin{figure}[H]
  \centering
  \includegraphics[width=0.72\textwidth]{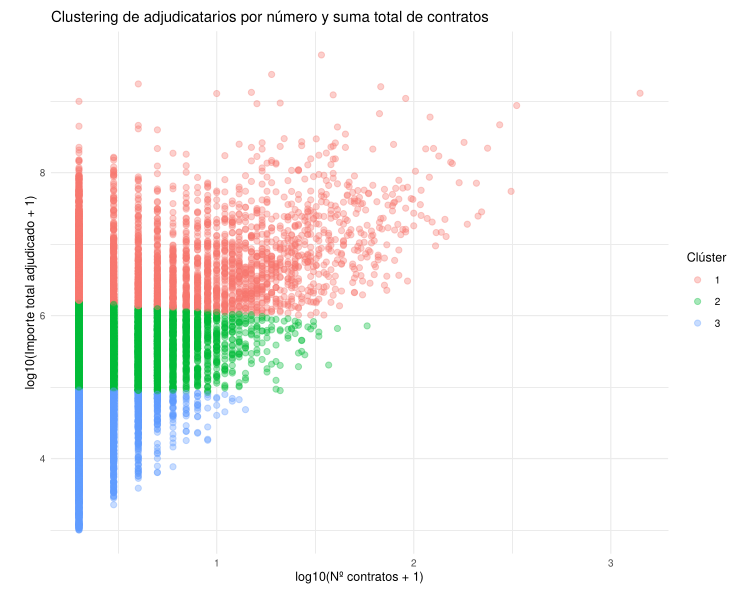}
  \caption{K-Means clustering of contractors by number of contracts
           (log10) and total awarded value (log10). Colours denote the
           three clusters: high-value operators (red), standard
           operators (green), and microoperators (blue).}
  \label{fig:clustering}
\end{figure}

The clustering reveals a pyramid structure: a comparatively small tier
of high-value operators concentrates high awarded amounts; a larger
intermediate tier represents the standard market; and the broadest
base aggregates occasional low-value participants.
Figure~\ref{fig:clustering} illustrates the three segments in
log-log space.

\subsection{Hypothesis Test: Works vs.\ Services}

A Wilcoxon--Mann--Whitney non-parametric test is applied to determine
whether awarded values differ significantly between Works ($n = 6{,}077$)
and Services ($n = 31{,}754$) contracts.

\textbf{Verification of parametric assumptions.}
Shapiro--Wilk tests applied to log-transformed values reject normality
for both groups ($p < 0.001$). Levene's test rejects equality of
variances ($F = 47.31$, $p < 10^{-12}$, $\text{df} = 1,\, 37{,}829$).
Both classical assumptions fail, justifying the non-parametric approach.

\textbf{Result.} $W = 134{,}427{,}022$, $p < 2.2 \times 10^{-16}$.
The null hypothesis of equal distributions is rejected.

\begin{table}[H]
\centering
\caption{Descriptive statistics by contract category.}
\label{tab:hypothesis}
\begin{tabular}{lrrrr}
\toprule
\textbf{Category} & \textbf{N} & \textbf{Median (€)} &
\textbf{Mean (€)} & \textbf{Max (€)} \\
\midrule
Works    & 6,077  & 296,266 & 2,616,298 & 446,614,679   \\
Services & 31,754 & 109,066 & 830,283   & 1,267,824,217 \\
\bottomrule
\end{tabular}
\end{table}

\begin{figure}[H]
  \centering
  \includegraphics[width=0.65\textwidth]{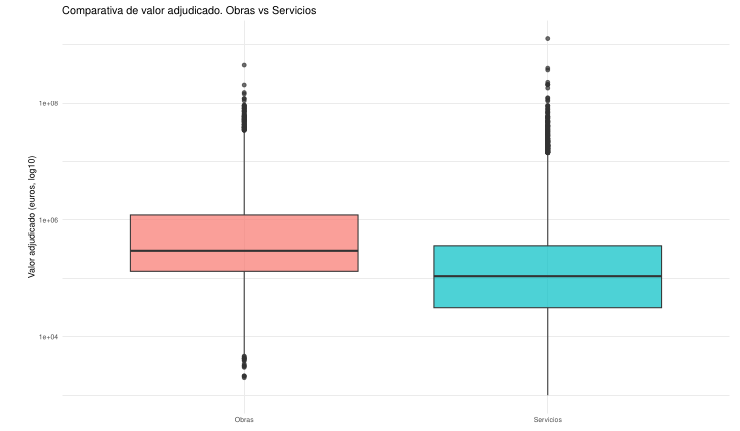}
  \caption{Boxplot of awarded values (log10) for Works and Services
           contracts. Works contracts present a substantially higher
           median and wider interquartile range.}
  \label{fig:works_services}
\end{figure}

Works contracts present significantly higher median and mean awarded
values, consistent with the capital-intensive nature of public
infrastructure investment. Figure~\ref{fig:works_services} confirms
this difference visually.

\section{Discussion and Limitations}

\textbf{Reproducibility.}
The full extraction pipeline and analytical R notebook are publicly
available at \url{https://github.com/mmunozpl/M2.851-May25}. The
dataset is published with a persistent DOI at
\url{https://doi.org/10.5281/zenodo.15120882} (CC0, 53~MB CSV) and
indexed in OpenAIRE, ensuring long-term accessibility.

\textbf{Reporting bias.}
The dataset captures only what the BOE publishes. Some contracts lack
awarded values, contractor names, or CPV codes. Publication delays may
place some contracts outside their actual award year, and completeness
varies across institutions.

\textbf{Classification errors.}
Complex HTML structures in certain announcement types lead to field
misclassification. The most notable artefact is the appearance of the
\textit{Tribunal Administrativo Central de Recursos Contractuales}
(TACRC) --- a dispute-resolution body, not a contractor --- in the
contractor rankings. This reflects a structural limitation of automated
extraction from semi-structured legal publications and is explicitly
documented as a known issue.

\textbf{Modelling scope.}
The linear model's low $R^2$ reflects the absence of structurally
imprescindible predictive variables: contract duration, number of
bidders, lot count, and technical complexity. These attributes are
either absent from BOE announcements or require substantially more
complex extraction logic. Non-linear approaches and additional
variables are identified as natural extensions.

\section{Conclusion}

We present a longitudinal, structured, and openly licensed dataset of
Spanish central-state public procurement covering 2014--2024, derived
from BOE Section~V-A through a fully automated and reproducible ETL
pipeline. The dataset comprises 97,154~observations across 15~variables
(60,456~contract award records after cleaning), is aligned with the
Open Contracting Data Standard, and is published under CC0 at
\url{https://doi.org/10.5281/zenodo.15120882}.

Exploratory analysis reveals strong value asymmetry, geographic and
sectoral concentration, and a structured three-tier market among
contractors. Multiple linear regression identifies geographic scope and
contract category as the principal structural determinants of awarded
value. K-Means clustering segments the contractor population into
three interpretable profiles. A Wilcoxon--Mann--Whitney test confirms
statistically significant and substantively large differences between
Works and Services contracts.

The dataset and pipeline constitute an open resource for public
procurement research, administrative data science, and evidence-based
policy evaluation in Spain.

\section*{Acknowledgements}

This work was developed as part of the Master's programme in Data
Science at the Universitat Oberta de Catalunya.

\section*{Data and Code Availability}

\textbf{Dataset (CC0):} Muñoz Plá, M. (2025). \textit{Licitaciones y
Contrataciones del BOE (2014--2024)} [Dataset]. Zenodo.
\url{https://doi.org/10.5281/zenodo.15120882}

\textbf{Code:} \url{https://github.com/mmunozpl/M2.851-May25}


\end{document}